\documentclass[aps,pra,twocolumn,groupedaddress]{revtex4}
\usepackage{graphicx}
\usepackage{dcolumn}
\usepackage{bm}
\usepackage{amsmath}
\usepackage{amssymb}
\usepackage{color}
\usepackage{slashed}

\begin{document}

\title{Arrival time from the general theory of quantum time distributions}

\author{Tajron Juri\'c}\email{tjuric@irb.hr}
\author{Hrvoje Nikoli\'c}\email{hnikolic@irb.hr}
\affiliation{Theoretical Physics Division, Rudjer Bo\v{s}kovi\'{c} Institute, P.O.B. 180, HR-10002 Zagreb, Croatia.}

%

\begin{abstract}
We further develop the general theory of quantum time distributions introduced in 
[D. Jurman and H. Nikoli\'c, Phys. Lett. A {\bf 396}, 127247 (2021)]
and apply it to find
the distribution of arrival times at the detector.
Even though the Hamiltonian in the absence of detector is hermitian,
the time evolution of the system before detection involves dealing 
with a non-hermitian operator obtained from the projection of the 
hermitian Hamiltonian onto the region in front of the detector.
Such a formalism eventually gives rise to a simple and physically sensible
analytical expression for the arrival time distribution, 
for arbitrary wave packet moving in one spatial dimension with negligible 
distortion.
\end{abstract}


\maketitle

\section{Introduction} 
 
In many quantum phenomena time is a random variable, in the sense that we do not know the exact time at which an event will happen. 
For instance, we do not know at which time an unstable atom will decay, so we describe it with
a probability density ${\cal P}(t)$, usually well described with an
exponential function of the form ${\cal P}(t)=\Gamma e^{-\Gamma t}$,
where $\Gamma$ is inversely proportional to the half-life time. But 
in standard quantum mechanics (QM) time is a classical variable, so why exactly is it random? 
And how to systematically find ${\cal P}(t)$ 
for {\em} any quantum phenomenon (not only for decays) in which $t$ behaves as a random variable?

A general systematic theory that answers such questions
for a large class of quantum phenomena has recently been developed in \cite{jur-nik}. 
It was found that for all such phenomena, ${\cal P}(t)$
can be written down in an exponential form 
\begin{equation}\label{exp}
 {\cal P}(t) = w(t) e^{-\int_0^t dt'w(t')} ,
\end{equation}
where $w(t)$ is a real non-negative function that can systematically be computed by the quantum theory, 
depending on details of the considered quantum system. Furthermore,
it was outlined how such a general theory can be applied not only to
decays, but also to problems such as arrival time, tunneling time and dwell time 
(for a review of these and other related problems with time in QM
see \cite{muga1,muga2}).

In this paper we apply the general theory developed in \cite{jur-nik}
to study in more detail the case of arrival time. 
(In \cite{jur-nik}, arrival time has been studied only for a  
special case of a rectangular wave packet.)
The arrival time 
problem has been attacked in the literature with many different 
approaches (see e.g. \cite{marchewka,wlodarz,
V1.10,V2.4,rovelli,delgado,galapon2,anastopoulos2,halliwell1,anastopoulos,vona,dhar,halliwell2,galapon,
leavens,durrtime1,durrtime2,maccone}
or the review \cite{muga-physrep} of older approaches).
Similarly to some other approaches (see e.g. \cite{allcock,echanobe} and references in \cite{jur-nik}), the
detection in our approach is described in terms of successive 
quantum collapses of the wave function. The 
main novelty in our approach, however, is a use of a new formalism based on a projected Hamiltonian.
In particular, while other approaches such as 
\cite{allcock,echanobe,muga99,halliwell08,halliwell10,kiukas} 
add an imaginary term to the Hamiltonian by hand, our projected Hamiltonian 
turns out to be non-hermitian automatically, due to a somewhat unexpected mathematical subtlety.
As we shall see, non-hermiticity arises from a careful investigation of the domains
of operators in the infinite-dimensional Hilbert space, where 
some naive formal manipulations, otherwise valid for finite-dimensional Hilbert spaces, 
turn out to be ill-defined in the infinite-dimensional case. 
Besides, most of the existing approaches to the arrival time do not seem 
to be a special case of a general systematic theory that can treat arrival time 
on an equal footing with all other phenomena, such as the decay time, in which time appears as a random variable.
(Some exceptions are \cite{V1.10,maccone} which are based on a general notion of a time operator
and \cite{anastopoulos} which has been extended 
to a general quantum temporal probability method for relativistic quantum field theory \cite{anastopoulos3}.)   
Those, we believe, are sufficient 
motivations to study in detail 
how arrival time can be described by the 
general systematic theory introduced in \cite{jur-nik}.

An additional motivation for this work arises from a recent argument 
by Das and Struyve \cite{das-struyve} that the theory \cite{jur-nik} and other 
similar approaches \cite{marchewka,wlodarz} to arrival time based on an exponential formula of the 
form (\ref{exp}) are not adequate because they lead to physically unacceptable 
exponential time-damping. In this paper we show that, when the general theory developed 
in \cite{jur-nik} is applied correctly, no exponential time-damping occurs.

The paper is organized as follows. In Sec.~\ref{SECoverview} we present
a brief overview of the general theory developed in \cite{jur-nik},
after which in Sec.~\ref{SECgen} we make some further extensions of the theory.
More specifically, the formalism as developed in \cite{jur-nik} only works for 
a certain class of pure initial states, while practical formulas 
in \cite{jur-nik} work only when the time interval $\delta t$
between subsequent measurements is sufficiently short. 
In Sec.~\ref{SECgen} we first extend the results of \cite{jur-nik} to arbitrary initial pure states,
then we further generalize the formalism to arbitrary mixed states, and finally 
find some practical formulas that do not depend on shortness of $\delta t$.
One of the formal tools in this theory is time evolution with a certain 
projected Hamiltonian, for which 
in Sec.~\ref{SECevol} we develop a formalism that specifies how to compute this time evolution in practice. 
Then in Sec.~\ref{SECarrival} we apply those results to analytically compute 
the arrival time distribution for a general wave packet traveling towards the detector,
in an approximation in which the distortion of the wave packet is neglected.
In Sec.~\ref{SECtrain} we study a special case of a Gaussian wave train
introduced in \cite{das-struyve}, to explain how the problem discussed there 
is resolved in our approach. Finally, the conclusions are drawn in Sec.~\ref{SECconcl}

\section{Overview of the general theory of time distributions}
\label{SECoverview}

Let us present a brief overview of the general theory of quantum time distributions developed in \cite{jur-nik}. 
Consider a quantum detector with only two possible outcomes: the detector either clicks or it doesn't.
We model it as an ideal projective measurement with projectors $\pi$ (corresponding to a click) and
$\bar{\pi}=1-\pi$ (corresponding to an absence of a click).
When the detector clicks, the state $|\psi\rangle$ of the measured system collapses as
\begin{equation}
 |\psi\rangle \rightarrow \frac{\pi |\psi\rangle}{||\pi |\psi\rangle||} .
\end{equation}
Likewise, as explained in more detail in \cite{jur-nik}, when the detector does not click, 
the state collapses as
\begin{equation}\label{ee3}
 |\psi\rangle \rightarrow \frac{\bar{\pi} |\psi\rangle}{||\bar{\pi} |\psi\rangle||} .
\end{equation}
The goal is to find the joint probability that the detector will click at time $t$ and not click
at times before $t$.

In the absence of measurement the state evolves according to $|\psi(t)\rangle=e^{-iHt}|\psi_0\rangle$
(unless stated otherwise, we work in units $\hbar=1$), 
where $H$ is the intrinsic Hamiltonian of the system (e.g.
the free Hamiltonian $H={\bf p}^2/2m$) and $|\psi_0\rangle$ is the initial state at the time $t_0=0$
satisfying 
\begin{equation}\label{ee4}
 \bar{\pi}|\psi_0\rangle=|\psi_0\rangle .
\end{equation}
To determine the time at which the click occurs,
we consider not one measurement but many measurements, each performed at a different time.
The time $\delta t$ between two subsequent measurements is short, so that the state $|\psi(t)\rangle$
does not change much during $\delta t$ by the intrinsic evolution governed by $H$.
So if the detector did not click at all $k$ subsequent measurements after the initial time, then
the state at time $t=k\delta t$ is
\begin{equation}\label{barcond}
 |\bar{\psi}_c(t)\rangle=\frac{V^k|\psi_0\rangle}{||V^k|\psi_0\rangle||} ,
\end{equation}
where $V\equiv \bar{\pi} e^{-iH\delta t}$. 
In the notation $|\bar{\psi}_c(t)\rangle$, the label $c$ denotes that it is the 
conditional state (describing the system {\em if} the detector did not click),
while the bar over $\psi$ denotes that this state satisfies 
$\bar{\pi}|\bar{\psi}_c(t)\rangle = |\bar{\psi}_c(t)\rangle$, i.e. 
lies in the projected Hilbert space $\bar{\pi}{\cal H}=\overline{\cal H}$, which is a subspace of the full Hilbert space ${\cal H}$. 
For small $\delta t$ and large $k$,
it was found in \cite{jur-nik} (for a different derivation see also Appendix \ref{APP})
that $V^k|\psi_0\rangle$ can be approximated with
\begin{equation}\label{ee6}
 V^k|\psi_0\rangle =e^{-i\overline{H}t} |\psi_0\rangle ,
\end{equation}
where 
\begin{equation}\label{overH}
 \overline{H}=\bar{\pi}H\bar{\pi} .
\end{equation}
Hence (\ref{barcond}) can be written as 
\begin{equation}\label{barcond2}
 |\bar{\psi}_c(t)\rangle=\frac{e^{-i\overline{H}t}|\psi_0\rangle}{||e^{-i\overline{H}t}|\psi_0\rangle||} .
\end{equation}
So if the detector did not click up to time $t-\delta t$, then the conditional probability that it will click at time $t$ is
\begin{equation}\label{p} 
 p(t)=\langle\psi_c(t)|\pi|\psi_c(t)\rangle ,
\end{equation}
where 
\begin{equation}\label{c2}
 |\psi_c(t)\rangle = e^{-iH\delta t}|\bar{\psi}_c(t-\delta t)\rangle .
\end{equation}
Therefore the probability that the detector will click at time $t$ is 
a product of two factors. One factor is the conditional probability (\ref{p}) above, and the other factor 
is the probability of this condition (that the detector will not click up to time $t-\delta t$)
itself.
This other factor, interesting in its own right, is explored in more detail in Appendix \ref{APP-wigner}. 
The final result for probability consisting of two factors,
expressed as a probability density ${\cal P}(t)$
with a continuous time, is shown in \cite{jur-nik} to be
\begin{equation}\label{Pdens}
 {\cal P}(t) = w(t) e^{-\int_0^t dt'w(t')} ,
\end{equation}  
where
\begin{equation}\label{w}
w(t)=\frac{p(t)}{\delta t}=\frac{1}{\delta t}\langle\psi_c(t)|\pi|\psi_c(t)\rangle .
\end{equation}

In addition, it was shown in \cite{jur-nik} that (\ref{Pdens}) is the solution of the integral equation
\begin{equation}\label{alt2}
 {\cal P}(t) = w(t) \left[ 1- \int_0^{t} dt'\, {\cal P}(t') \right] .
\end{equation}
This integral equation will turn out to be a very useful calculation trick
that saves a lot of work by avoiding explicit computation of the exponential factor in (\ref{Pdens}),
as will be demonstrated in Sec.~\ref{SECarrival}.

\section{Further extensions of general theory}
\label{SECgen}

In this section we make some extensions of the results of Sec.~\ref{SECoverview}
that have not been obtained in \cite{jur-nik}.

\subsection{Extension to arbitrary pure states}

First, note that
Eq.~(\ref{ee6}) is only valid for states which satisfy (\ref{ee4}). For general pure states this 
generalizes to
\begin{equation}\label{ee6gen}
 V^k|\psi\rangle =e^{-i\overline{H}t} \bar{\pi}|\psi\rangle ,
\end{equation}
in the limit $\delta t\to 0$, $k\to\infty$, $k\delta t=t$ finite.
To prove this, we first decompose $|\psi\rangle$ as 
\begin{equation}\label{decomp}
|\psi\rangle=\bar{\pi}|\psi\rangle+\pi |\psi\rangle ,
\end{equation}
so that the first term $\bar{\pi}|\psi\rangle$ is in the projected Hilbert space 
$\bar{\pi}{\cal H}$ and the second term $\pi|\psi\rangle$
is in its complement $\pi{\cal H}$.
Hence the first term 
in (\ref{decomp})
satisfies (\ref{ee6gen}) due to (\ref{ee6}), so it only
remains to prove that the second term 
in (\ref{decomp})
satisfies (\ref{ee6gen}).
But the right-hand side of (\ref{ee6gen}) applied to the second term $\pi|\psi\rangle$ 
in (\ref{decomp})
gives zero,
so we only need to show that the left-hand side of (\ref{ee6gen}) applied to the second term 
in (\ref{decomp})
also gives zero.
Therefore
we consider $V^k\pi|\psi\rangle=V^{k-1}V\pi|\psi\rangle$ in the limit above.
But $V=\bar{\pi} e^{-iH\delta t}$, so in the limit we have 
$V\pi|\psi\rangle \to \bar{\pi}\pi|\psi\rangle=0$, implying $V^k\pi|\psi\rangle \to 0$. Thus the left-hand side of (\ref{ee6gen})
is zero when applied to the second term in (\ref{decomp}),
which proves (\ref{ee6gen}). 
A different derivation of (\ref{ee6gen}) is presented also in Appendix \ref{APP}.

\subsection{Extension to mixed states}

Second, the theory can be generalized to include the mixed states.
For instance, (\ref{ee3}) generalizes to
\begin{equation}\label{ee3g}
 \rho \rightarrow \frac{\bar{\pi}\rho\bar{\pi}}{{\rm Tr}(\bar{\pi}\rho\bar{\pi})} ,
\end{equation}
where $\rho$ is an arbitrary pure or mixed state, generalizing the case of a pure state $\rho=|\psi\rangle\langle\psi|$.
Similarly, (\ref{barcond}) and (\ref{barcond2}) generalize to
\begin{eqnarray}\label{barcondg}
 \bar{\rho}_c(t) &=& \frac{V^k\rho_0(V^k)^{\dagger}}{{\rm Tr}(V^k\rho_0(V^k)^{\dagger})} 
\nonumber \\
&=& \frac{e^{-i\overline{H}t} \bar{\pi}\rho_0\bar{\pi}(e^{-i\overline{H}t})^{\dagger}}
{{\rm Tr}(e^{-i\overline{H}t} \bar{\pi}\rho_0\bar{\pi}(e^{-i\overline{H}t})^{\dagger})} ,
\end{eqnarray}
where the generalization (\ref{ee6gen}) is also taken into account.
Likewise, (\ref{c2}) and (\ref{w}) generalize to
\begin{equation}\label{c2gen}
 \rho_c(t) = e^{-iH\delta t} \bar{\rho}_c(t-\delta t) e^{iH\delta t} ,
\end{equation}
\begin{equation}\label{wgen}
w(t)=\frac{1}{\delta t} {\rm Tr}(\rho_c(t)\pi) .
\end{equation}
With these generalizations, the probability density is again given by (\ref{Pdens}). 
Note that this probability density can also be expressed in terms of a POVM that has a nonlinear dependence on the 
quantum state \cite{jur-nik}. 

\subsection{Extension to arbitrary $\delta t$}

Finally we show that, under a certain condition, 
a useful expression for $V^k$ can be derived {\em without} the limit
$\delta t\to 0$, $k\to\infty$. In Sec.~\ref{SECevol} we shall see that a physically interesting projector 
and Hamiltonian formally satisfy $\bar{\pi}H\bar{\pi}=\bar{\pi}H$. This motivates us to study an operator $\bar{\xi}$
that satisfies the condition
\begin{equation}\label{magic0}
\bar{\xi}H\bar{\xi}=\bar{\xi}H .
\end{equation}
Here $\bar{\xi}$ does not need to be a projector, it only needs to be an operator that satisfies (\ref{magic0}).
Physically, it can be interpreted as a Kraus operator associated with a POVM measurement that does not need 
to be a projective measurement.

We start from the obvious identity
\begin{equation}
\bar{\xi} H^n = \bar{\xi} H H^{n-1} ,
\end{equation}
which can also be rewritten as
\begin{equation}
\bar{\xi} H^n \stackrel{(\ref{magic0})}{=}  \bar{\xi} H \bar{\xi} H^{n-1} =
\bar{\xi} H \bar{\xi} H H^{n-2} = (\bar{\xi} H)^2 H^{n-2}.
\end{equation}
Thus, by induction, we see that in general 
\begin{equation}
 \bar{\xi} H^n = (\bar{\xi} H)^m H^{n-m} , 
\end{equation}
for $0\leq m\leq n$. In particular, for $m=n$ we have a very useful identity
\begin{equation}\label{ex2}
 \bar{\xi} H^n = (\bar{\xi} H)^n . 
\end{equation}
This identity implies
\begin{eqnarray}\label{ex3}
 \bar{\xi} e^{-iHt} &=& \sum_n \frac{(-it)^n}{n!} \bar{\xi} H^n
\nonumber \\
 & \stackrel{(\ref{ex2})}{=} & \sum_n \frac{(-it)^n}{n!} (\bar{\xi} H)^n = e^{-i\bar{\xi}Ht} .
\end{eqnarray}
Defining $V=\bar{\xi} e^{-iH\delta t}$, (\ref{ex3}) implies
\begin{equation}\label{ex4}
 V=e^{-i\bar{\xi}H\delta t} .
\end{equation}
Hence
\begin{equation}\label{ex5}
 V^k \stackrel{(\ref{ex4})}{=} e^{-i\bar{\xi}Hk\delta t}
\stackrel{t\equiv k\delta t}{=} e^{-i\bar{\xi}Ht} .
\end{equation}
From (\ref{ex5}) and (\ref{ex3}) we then get the final result useful for practical applications
\begin{equation}\label{ex6}
 V^k = \bar{\xi} e^{-iHt}. 
\end{equation}
It is useful because it only involves evolution with the ordinary Hamiltonian $H$,
the evolution by which is supposed to be well understood. We repeat that the equations with $\bar{\xi}$ above 
are exact for arbitrary $\delta t$ and $k$, i.e. we have not used the $\delta t\to 0$, $k\to\infty$ limit.

\section{Evolution with the projected Hamiltonian}
\label{SECevol}

To apply the general theory outlined in Sec.~\ref{SECoverview}, in (\ref{barcond2}) one has to compute the time evolution
governed by the projected Hamiltonian $\overline{H}$ defined by (\ref{overH}). 
In \cite{jur-nik} it was assumed that it can be done, but an explicit method for doing it has not been developed.
In this section we explicitly develop a method for computing the time evolution with $\overline{H}$,
applicable to a certain relatively large class of problems.

The crucial issue turns out to be the hermiticity of $\overline{H}$. Naively it seems obviously hermitian, 
because $H$ and $\bar{\pi}$ are hermitian so 
$\overline{H}^{\dagger}=\bar{\pi}^{\dagger}H^{\dagger}\bar{\pi}^{\dagger}=\bar{\pi}H\bar{\pi}=\overline{H}$.
In \cite{jur-nik} it was indeed assumed that $\overline{H}$ is hermitian, which implied that 
$e^{-i\overline{H}t}$ was unitary. Consequently, the denominator in (\ref{barcond2}) has been dropped in \cite{jur-nik} 
because it was assumed that it is equal to 1. Indeed, in a Hilbert space of a finite dimension, 
such a reasoning is perfectly correct. However, in an infinite dimensional Hilbert space one has to be 
much more careful. If one takes the rules valid for finite dimensional Hilbert spaces and naively applies them 
to infinite dimensional ones, one may encounter various inconsistencies and apparent paradoxes \cite{gieres}. 

First we have to be aware that when considering wave functions as square-integrable functions,
we are dealing with an infinite dimensional Hilbert space usually denoted by $L^2(\mathcal{M})$, 
where $\mathcal{M}$ is the underlying space manifold. Operators in such a Hilbert space, 
that is linear mappings $A:L^2(\mathcal{M})\longrightarrow L^2(\mathcal{M})$, 
are a bit more complicated and subtle objects 
than 
matrices (operators on a finite dimensional Hilbert space). Since all of the relevant operators in QM 
(like Hamiltonian, momentum, position, etc.) are usually certain differential operators, 
one needs to restrict the possible types of ``allowable'' wave functions (for example  
differentiable functions $C^{1}(\mathcal{M})$, twice differentiable functions 
$C^{2}(\mathcal{M})$, smooth functions $C^{\infty}(\mathcal{M})$ or 
compactly supported smooth functions $C^{\infty}_{\rm c}(\mathcal{M})$), 
that is elements of $L^2(\mathcal{M})$ such that the actions of these operators are well defined. 
Also, when dealing with differential equations like Schr\"odinger equation or an eigenvalue equation 
for any differential operator, one often needs to introduce some boundary conditions. 
All this leads to a conclusion that a properly defined operator $A$ on $L^2(\mathcal{M})$ 
comes with its domain $\mathcal{D}(A)$, which is at best a dense 
subset \footnote{We will always assume that $\mathcal{D}(A)$ is dense in $L^2(\mathcal{M})$, 
that is $\overline{\mathcal{D}(A)}=L^2(\mathcal{M})$, meaning that the topological closure of 
$\mathcal{D}(A)$ is equal to $L^2(\mathcal{M})$. Namely, $\mathcal{D}(A)$ is usually not  a 
complete space, so dense means that the space together with limit vectors of all of its Cauchy sequences 
is exactly equal to $L^2(\mathcal{M})$.} 
of $L^2(\mathcal{M})$ (for more details see \cite{gitman,juric}). 
Having this in mind, the hermiticity condition \footnote{The term ``hermitian operator'' is often abused in physics 
literature. Sometimes it means (\ref{hermcon}), which mathematicians call symmetric operator, while sometimes 
it means a self-adjoint operator. A self-adjoint operator $A$ is a symmetric operator for which the domain of 
its adjoint $A^\dagger$ satisfies $\mathcal{D}(A^\dagger)=\mathcal{D}(A)$. On finite dimensional Hilbert spaces 
the three notions coincide \cite{juric}.}, 
which in finite dimensional case was $A=A^\dagger$, now reads
\begin{equation}\label{hermcon}
(\psi,A\varphi)=(A\psi,\varphi), \quad \forall \psi,\varphi\in\mathcal{D}(A)\subset L^2(\mathcal{M}) ,
\end{equation}
where 
\begin{equation}
(\psi,\varphi)=\int_{\mathcal{M}} d\mu(x)\,
\psi^*(x)\varphi(x), \quad \forall \psi,\varphi\in L^2(\mathcal{M})
\end{equation}
is the usual sesquilinear inner product on $L^2(\mathcal{M})$ and $d\mu(x)$ is the Lebesgue measure. 

Let us now go back to our projected Hamiltonian $\overline{H}$.  Since it is given by the 
composition $\overline{H}=\bar{\pi}H\bar{\pi}$ 
of the intrinsic Hamiltonian $H$ and projectors $\bar{\pi}$, 
in order to properly define $\overline{H}$ with the right domain we first present all the operators 
and their domains in the so called coordinate representation. 
For simplicity we take $H$ to be
the free Hamiltonian of a particle moving on a real line, so in the coordinate representation we have  
\begin{equation}\label{fH}
H=-\frac{1}{2m}\frac{d^2}{dx^2}, \quad \mathcal{D}(H)\subset L^2(\mathcal{M}) 
\end{equation}
and $\mathcal{M}=\mathbb{R}$.
The projectors that will be of physical interest have the coordinate representation of the form
\begin{equation}\label{pc}
 \bar{\pi}=\chi_D(x) ,
\end{equation}
where the domain $D$ is a subspace of $\mathcal{M}$ and
\begin{equation}\label{pc2}
\chi_D(x)\equiv\left\{ 
\begin{array}{ll}
 \displaystyle 1 & \;\; {\rm for} \;\; x\in D  \\
 0 & \;\; {\rm for} \;\; x\notin D 
\end{array}
\right.
\end{equation}
is the characteristic function associated with $D$. 
Such projectors $\bar{\pi}$ can be abstractly represented (at least up to a boundary term) as
\begin{equation}
 \bar{\pi}=\int_D dx |x\rangle\langle x| ,
\end{equation}
where $|x\rangle$ and $\langle x|$ are the Dirac's bra and ket representation of the eigenfunctions of the position operator $x$. 
In order for the composition $\bar{\pi}H\bar{\pi}$ to be well defined, the domain of $\overline{H}$ is given by
\begin{equation}\label{presjek}
\mathcal{D}(\overline{H})=\mathcal{D}(H)\cap\mathcal{D}(\bar{\pi})=\mathcal{D}(H) ,
\end{equation}
since $\bar{\pi}$ is a bounded operator and $\mathcal{D}(\bar{\pi})=L^2(\mathcal{M})$ is insured by the 
BLT theorem \footnote{The BLT (Bounded Linear Transformation) theorem states that any bounded 
and densely defined linear map 
$A: \mathcal{D}(A)\subset X\longrightarrow Y$ has a unique extension $\tilde{A}\supset A: X\longrightarrow Y$ 
such that $\tilde{A}$ is bounded, linear and with the same operator norm $\|\tilde{A}\|=\left\|A\right\|$. 
Here $X$ has to be a normed space and $Y$ a complete normed space (Banach space). In practice this means that  a 
bounded operator $A$ defined on some dense domain $\mathcal{D}(A)\subset X$ can always be represented uniquely by 
a bounded operator $\tilde{A}$ that act on the whole space $X$, that is the completion of the domain $\overline{\mathcal{D}(A)}=X$. 
In our paper we always have $X=Y=L^2(\mathcal{M})$ and therefore all the bounded operators (such as $\pi$ or $\bar{\pi}$) 
are actually the unique extensions, so their domain is the whole Hilbert space $L^2(\mathcal{M})$. For more details on 
bounded operators see \cite{reed}.}. 
Now the projected Hamiltonian in the coordinate representation
is determined by elementary calculus
\begin{eqnarray}\label{pravi}
\overline{H}\psi(x) &=& (\bar{\pi}H\bar{\pi}\psi)(x)=-\frac{1}{2m}\chi_{D}(x)\left[\chi_{D}(x)\psi(x)\right]''
\nonumber \\
&=&
-\frac{1}{2m}
\chi_{D}\left[ \chi_{D}''\psi+2\chi_{D}'\psi'+\chi_{D}\psi''\right] ,
\end{eqnarray}
where we used \eqref{fH} and \eqref{pc},
the primes denote derivatives over $x$, and in the last line we suppressed explicit 
writing of the dependence on $x$.
Note that $\chi_{D}'(x)$ and $\chi_{D}''(x)$ are zero everywhere, except at the boundary of $D$.
Hence particularly interesting is the case when the domain $D$ is an open set, because 
then $D$ does not contain
its boundary so \eqref{pc2} vanishes on the boundary. So when $D$ is open, then $\chi_{D} \chi_{D}''$ and $\chi_{D} \chi_{D}'$
vanish everywhere, which implies that \eqref{pravi} enormously simplifies to   
\begin{equation}\label{pravi2}
\overline{H}\psi(x)=\chi_D(x)H\psi(x)= \left\{ 
\begin{array}{ll}
 \displaystyle H\psi(x) & \;\; {\rm for} \;\; x\in D  \\
 0 & \;\; {\rm for} \;\; x\notin D .
\end{array}
\right.
\end{equation}

Now we are ready to test the hermiticity \eqref{hermcon} of the projected Hamiltonian 
$\overline{H}$ \eqref{pravi}. Using \eqref{pravi}, testing the condition \eqref{hermcon} in general boils down to testing whether 
\begin{equation}\label{non-herm1}
\left[ \int_D dx\, \varphi^*(x) H \psi(x) \right]^* = 
 \int_D dx\, \psi^*(x)H \varphi(x)
\end{equation}
is valid $\forall \psi,\varphi \in \mathcal{D}(H)$.
The validity of this highly depends on the choice for the domain of the free Hamiltonian 
$\mathcal{D}(H)$ and the choice of the subspace $D$. Here the underlying space is the whole real line,
$\mathcal{M}=\mathbb{R}$, so we are working with the Hilbert space $L^2(\mathbb{R})$ and the domain of a hermitian 
(even more, self-adjoint!) free Hamiltonian $H$ is given by the space of 
Schwartz functions \footnote{The space of Schwartz functions, or just Schwartz space $\mathcal{S}(\mathbb{R})$, 
is a vector space that contains all functions whose derivatives (multiplied by any monomial in $x$) are
rapidly decreasing \cite{reed}. These functions are often called \textit{test functions}.}, 
while the subspace $D$ is an interval in $\mathbb{R}$. Hence,
after a partial integration,
it is easy to see that the condition \eqref{non-herm1} is satisfied up to a boundary term. Since we are working with the domain that 
are Schwartz functions, that only decay rapidly in the infinity and therefore take an arbitrary value on the boundary 
of the interval $D$, there are uncountably infinitely many  functions $\psi$ and $\varphi$ in 
$\mathcal{D}(H)=\mathcal{S}(\mathbb{R})$ such that \eqref{non-herm1} is not fulfilled. This implies that the 
projected Hamiltonian $\overline{H}$ is a {\em non-hermitian} operator on the Hilbert space $L^2(\mathbb{R})$.

Indeed, from a physical point of view, it is very important and desirable for $\overline{H}$ to be non-hermitian. 
If $\overline{H}$ was hermitian then $e^{-i\overline{H}t}$ would be unitary, so
the denominator in (\ref{barcond2}) would be equal to one and hence time-independent.
The denominator in (\ref{barcond2}) originates from the change of normalization induced
by collapse (\ref{ee3}), so with a hermitian $\overline{H}$ one could not account for the needed
change of normalization. As we shall discuss in more detail in Sec.~\ref{SECtrain}, 
the time-dependent normalization implied by non-hermiticity resolves the problem raised in  \cite{das-struyve}.

Note that, mathematically, one could redefine $\overline{H}$ to be a hermitian operator, 
but in that case one would need to change its domain to a certain subspace, 
namely $\tilde{\mathcal{D}}(\overline{H})\subsetneq \mathcal{D}(H)$.  
To provide the hermiticity of this new $\overline{H}$, the domain $\tilde{\mathcal{D}}(\overline{H})$ would have to
consist of wave functions that satisfy certain  boundary conditions 
on the boundary of $D$, so that the boundary term appearing after the partial 
integration vanishes.
However, the boundary conditions would have physical consequences, 
e.g. a reflection of the wave packet impinging on the boundary. Such a reflection might have a sensible physical interpretation
if the boundary represented a physical wall,
but in our case there is no physical wall. Physically, the projection with $\bar{\pi}$ originates
from wave function collapse associated with the {\it absence} of a click, i.e. with gaining information that the particle is {\em not} positioned outside of $D$. Physically, 
such a projection should play a rather passive role on the wave function,
in the sense that it should only dismiss a part of the wave function 
(the part outside of $D$)
which is irrelevant, without affecting the relevant part that is not dismissed (the part in $D$).  
Hence the boundary itself should not have a physical effect on the wave function.

The physical requirement that the boundary should not have a physical effect has one additional consequence.
In (\ref{pravi}) we have seen that the action of $\overline{H}$ creates nontrivial boundary effects,
unless $D$ is open, in which case the boundary effects vanish in (\ref{pravi2}).
This means that, 
for the projector $\bar{\pi}$ associated with an absence of a click, the domain 
$D$ is always interpreted as an {\em open} domain. 

Now we are ready to compute the time evolution governed by $\overline{H}$. This can be done by computing
$\bar{\phi}(x,t)\equiv\langle x|e^{-i\overline{H}t} |\psi_0\rangle$ satisfying the pseudo-Schr\"odinger equation
\begin{equation}\label{pseudosch}
 \overline{H}\bar{\phi}(x,t)=i\partial_t\bar{\phi}(x,t) ,
\end{equation}
where $\overline{H}$ is defined by (\ref{pravi2}).
Hence the appropriate solution of (\ref{pseudosch}) can be written as
\begin{equation}\label{barchi}
 \bar{\phi}(x,t)=\chi_D(x)\psi(x,t) ,
\end{equation}
where $\psi(x,t)$ is the solution of the true Schr\"odinger equation
\begin{equation}\label{sch}
 H\psi(x,t)=i\partial_t \psi(x,t) 
\end{equation}
describing the system in the absence of measurement.   
(Note that (\ref{barchi}) with (\ref{sch}) can formally be written as 
\begin{equation}
|\bar{\phi}(t)\rangle=\bar{\pi}e^{-iHt} |\psi(0)\rangle, 
\end{equation}
with the evolution operator 
$\bar{\pi}e^{-iHt}$ having the same form as (\ref{ex6}).)
Hence (\ref{barcond2}) implies that $\bar{\psi}_c(x,t)=\langle x|\bar{\psi}_c(t)\rangle$ is given by
\begin{equation}\label{barpsix}
 \bar{\psi}_c(x,t)=\displaystyle\frac{\chi_D(x)\psi(x,t)}{\sqrt{\int_D dx'\, |\psi(x',t)|^2}} .
\end{equation}

In the analysis above, the domain $D\subset \mathbb{R}$ was still quite arbitrary.
Its exact specification is determined by the concrete physical problem.
In particular, for a study of the arrival time in the next section, we are interested in the following
configuration. The initial wave function $\psi(x,0)$ is a wave packet with a support on the left region $D_L=(-\infty,0)$. 
This wave packet 
travels from left to the right towards the detector at $x=0$. Assuming that the detector has the width $l$, the detector 
region is characterized by the closed domain $D_{\rm det}=[0,l]$. Finally, on the right from the detector we have the right domain
$D_R=(l,\infty)$. Hence the projector associated with the detector is
\begin{equation}
 \pi=\int_{D_{\rm det}} dx\, |x\rangle\langle x|,
\end{equation}
while $\bar{\pi}$ is defined as
\begin{equation}
 \bar{\pi} = \bar{\pi}_L+\bar{\pi}_R=
\int_{D_L} dx\, |x\rangle\langle x| + \int_{D_R} dx\, |x\rangle\langle x| .
\end{equation}
We have chosen $D_{\rm det}$ to be closed because $D_L$ and $D_R$ are open and we must have 
$\bar{\pi}+\pi=1$.
Since the conditional wave function $\bar{\psi}_c(x,t)$ is initially in $D_L$ and never
enters $D_{\rm det}$, it follows that it cannot propagate to $D_R$. Hence, by analysis similar to the analysis that led to
(\ref{barpsix}), one finds
\begin{equation}\label{barpsix2}
 \bar{\psi}_c(x,t)=\displaystyle\frac{\chi_{D_L}(x)\psi(x,t)}{\sqrt{\int_{D_L} dx'\, |\psi(x',t)|^2}} .
\end{equation}

\section{Arrival time distribution}
\label{SECarrival}

Now we apply the results of Sec.~\ref{SECoverview} and of the last paragraph of Sec.~\ref{SECevol} 
to compute the arrival time distribution. The central quantity that needs to be computed is (\ref{w})
given by
\begin{equation}\label{w2}
w(t)=\frac{1}{\delta t}\int_0^l dx\, \rho_c(x,t) ,
\end{equation} 
where $\rho_c(x,t)=|\psi_c(x,t)|^2$. In the absence of measurement, the wave function $\psi(x,t)$ is a wave packet that 
propagates from the left to the right. For the free Hamiltonian $H=p^2/2m$, the wave packet has a constant average
velocity $v$. (By the Ehrenfest theorem, the average velocity is equal to $\langle p\rangle/m$, where $\langle p\rangle$ is the usual 
quantum mechanical average of the momentum.) From (\ref{c2}) we see that $\psi_c(x,t)$ enters the detector region 
$D_{\rm det}=[0,l]$ only during a short time from $t-\delta t$ to $t$. During this time $\psi_c(x,t)$ penetrates 
only a small distance $\delta l\simeq v\delta t < l$ of the detector region, so (\ref{w2}) can be approximated with
\begin{equation}\label{w3}
w(t)=\frac{1}{\delta t}\int_0^{\delta l} dx\, \rho_c(x,t) \simeq \frac{\delta l}{\delta t} \rho_c(0,t) \simeq
v\rho_c(0,t).
\end{equation}
This is quite remarkable from a physical point of view because it eliminates the 
dependence on the detector parameters $\delta t$ (the time resolution of the detector)
and $l$ (the width of the detector), by replacing them with a dependence on the 
(detector-independent) velocity $v$ of the wave packet.

Next we want to relate (\ref{w3}) to $\bar{\rho}_c(x,t)=|\bar{\psi}_c(x,t)|^2$, but since $\bar{\rho}_c(0,t)=0$, 
we consider 
\begin{equation}
 \bar{\rho}_c(0^-,t)=\lim_{\varepsilon\to 0^+}\bar{\rho}_c(-\varepsilon,t) ,
\end{equation}
which obeys $\bar{\rho}_c(0^-,t)=\rho_c(0,t)$. Hence (\ref{w3}) can be written as
\begin{equation}\label{w4}
w(t)=v\bar{\rho}_c(0^-,t).
\end{equation}
Using (\ref{barpsix2}), this can be expressed in terms of the probability density 
$\rho(x,t)=|\psi(x,t)|^2$ in the absence of measurement, as
\begin{equation}\label{w5}
w(t)=\frac{v\rho(0,t)}{\int_{-\infty}^0 dx\, \rho(x,t)} .
\end{equation}

Now, assuming that the wave function $\psi(x,t)$ in the absence of measurement is known,
it remains to use the result (\ref{w5}) to compute the arrival time probability density (\ref{Pdens}).
In principle one should compute the integrals over $x$ in (\ref{w5}) and over $t'$ in (\ref{Pdens}),
which is straightforward but complicated. Fortunately there is a neat shortcut, based on using the integral equation (\ref{alt2}), 
that completely avoids 
the explicit computation of the integrals.
Since the wave packet travels to the right without a reflection, the particle will eventually be detected with certainty,
so 
\begin{equation}\label{norm1}
\int_0^{\infty} dt'\, {\cal P}(t')=1 .
\end{equation}
Hence (\ref{alt2}) can be written as
\begin{equation}\label{alt3}
w(t)=\frac{{\cal P}(t)}{\int_t^{\infty}dt'\, {\cal P}(t')} .
\end{equation}
This looks very similar to (\ref{w5}), which is not a coincidence. 
The denominator in (\ref{w5}) is the probability that the particle at time $t$ is still in region 
in front of the detector,
the denominator in (\ref{alt3}) is the probability that the particle will be detected after the time $t$,
and physically it is quite intuitive to expect that those two probabilities should be the same. 
If that expectation is true then the numerators in (\ref{w5}) in (\ref{alt3}) must also be the same,
so one expects
\begin{equation}\label{arrival_heur}
 {\cal P}(t) = v\rho(0,t) .
\end{equation}

The expectation (\ref{arrival_heur}) can be confirmed mathematically by exploiting the mathematical 
similarity between (\ref{alt3}) and (\ref{w5}). For that purpose
we neglect the distortion of the wave packet during its time evolution, i.e. we use the approximation
\begin{equation}\label{shift0}
 \rho(x,t)=\rho(x-vt,0) . 
\end{equation}
(In fact, for relativistic massless particles such as photon this relation can be exactly satisfied 
because any sufficiently differentiable function of the form $\psi(x-ct)$ satisfies
the relativistic wave equation $[c^{-2}\partial_t^2-\partial_x^2]\psi(x,t)=0$.)  
Thus, for instance, we also have
\begin{equation}\label{shift1}
 \rho(x,t)=\rho(0,t-x/v) ,
\end{equation}
or more generally
\begin{equation}\label{shift2}
 \rho(x,t)=\rho(x-a,t-a/v) 
\end{equation} 
for an arbitrary spatial shift $a$. Hence the denominator in (\ref{w5}) can be written as
\begin{equation}\label{transform}
\int_{-\infty}^0 dx\, \rho(x,t) = \int_{-\infty}^0 dx\, \rho(0,t-x/v)
= \int_{t}^{\infty} dt'\, v\rho(0,t') ,
\end{equation}
where in the first equality we used (\ref{shift1}) (or (\ref{shift2}) with $a=x$),
while in the second equality we introduced a new integration variable $t'=t-x/v$, so that
$dt'=-dx/v$. 
Hence (\ref{w5}) becomes 
\begin{equation}\label{w6}
w(t)=\frac{v\rho(0,t)}{\int_{t}^{\infty} dt'\, v\rho(0,t')} .
\end{equation}
Comparing it with (\ref{alt3}), we see immediately that
\begin{equation}\label{Pfinal}
  {\cal P}(t)=v\rho(0,t) ,
\end{equation}
which confirms the expectation (\ref{arrival_heur}).
More precisely, the comparison of (\ref{w6}) with (\ref{alt3}) only implies that ${\cal P}(t)=\alpha v\rho(0,t)$,
where $\alpha$ is some overall normalization constant, but it is easy to check that (\ref{Pfinal}) is correctly 
normalized, i.e. that $\alpha=1$. This is seen from
\begin{equation}\label{norm1.1}
\int_0^{\infty} dt'\, {\cal P}(t')=\int_0^{\infty} dt'\, v\rho(0,t')=\int_{-\infty}^0 dx\, \rho(x,0)=1 ,
\end{equation}
which is consistent with (\ref{norm1}). In the second equality we used (\ref{transform}) with $t=0$,
while in the last equality we used the fact that the initial wave function $\psi(x,0)$ is correctly 
normalized and has a support in $D_L=(-\infty,0)$. Hence (\ref{Pfinal}) is our final general result 
for the time distribution of arrival times, which indeed is very simple and physically rather intuitive.
It is just the average flux of particles at the position of the detector at $x=0$,
that is, the position probability density at the position of the detector times the average velocity of the particles.

Note that (\ref{Pdens}) contains an exponential time-damping factor,
but (\ref{Pfinal}) does not contain such a factor. It can be traced back to the fact that (\ref{barcond2}) contains a time-dependent 
normalization factor that exactly cancels the exponential 
time-damping in (\ref{Pdens}). This is a generalization of the result
in \cite{jur-nik}, where such a cancellation has been obtained 
for a rectangular wave packet by an explicit computation 
of the integrals, without using the integral equation (\ref{alt2}).
Fundamentally, the cancellation of the normalization factor with the exponential time damping is a 
direct consequence of (\ref{wig-up}) in Appendix \ref{APP-wigner}, 
where the derivation and probabilistic meaning of (\ref{wig-up}) is presented in detail.

Finally, we repeat that our analytic result 
(\ref{Pfinal})
is based on two approximations, 
both of which have limitations.
One approximation is neglecting the distortion of the wave packet in (\ref{shift0}).
For instance, for the free Gaussian wave packet, the distortion is described 
by the time-dependent width \cite{cohen}
\begin{equation}
\sigma(t)=\sigma_0\sqrt{1+\frac{t^2}{\tau^2}} ,
\end{equation}
where $\sigma_0$ is the initial width and 
\begin{equation}
\tau=\frac{2m\sigma_0^2}{\hbar} ,
\end{equation}
in physical units with $\hbar\neq 1$.
Hence the distortion is negligible during the time $t\ll\tau$, which in experiment can be
set up to be long enough by preparing the initial state with a sufficiently large $\sigma_0$.  
The other approximation is (\ref{w3}); 
the propagation of a realistic wave packet in the detector region
is not given simply by the average velocity $v$ of the free wave packet.
In a more realistic analysis $\delta l$ would be interpreted as the length that must be penetrated
before a signal is recorded, so $\delta l/\delta t$ would still be interpreted as a parameter of the detector.
A more realistic analysis that avoids both inaccuracies would require careful numerical
computations, which are planed to be studied in a separate work. However, in a numerical analysis
one must choose a specific wave packet, so in this way one cannot obtain a general result.
A benefit of our approximations is that they enable to obtain a general analytic result
referring to a wave packet of an arbitrary shape. We believe that it contributes to conceptual
qualitative understanding, even if it is not reliable for a precise comparison of theoretical predictions with 
actual measurements.  
We also stress that the cancellation discussed in the previous paragraph does not depend on those approximations,
because the cancellation is a general consequence of Eq.~(\ref{wig-up}), which only depends on the assumption that
$\delta t$ is sufficiently small.

\section{Gaussian wave train}
\label{SECtrain}

Now let us apply the final general result (\ref{Pfinal}) to the Gaussian wave train studied by Das and Struyve 
in \cite{das-struyve}. They studied it with a motivation to 
demonstrate that theories of arrival time based on a formula 
of the form of (\ref{Pdens}) lead to exponential 
time-damping, which in the case of a Gaussian wave train 
seems physically inadequate. 
However, as discussed in the 
next to 
last paragraph of 
Sec.~\ref{SECarrival}, it is clear that there is no such damping
in our theory. Let us discuss it in more detail.

Let $\psi_G(x,t)$ be a normalized Gaussian wave packet moving to the right with the velocity $v$, 
defined so that its maximum is at $x=0$ for $t=0$. Then $\psi_G(x+a,t)$ is a similar Gaussian wave packet 
with the maximum at $x=-a$ for $t=0$.
The Gaussian wave train is a wave function of the form
\begin{equation}\label{train}
 \psi(x,t)=\frac{1}{\sqrt{N}}\sum_{n=1}^N \psi_G(x+a_n,t) .
\end{equation}
Here we assume that all $a_n$ are positive and much larger
than the width of the Gaussian packet, so that 
initially the support of (\ref{train}) is negligible outside of $D_L$.
We also assume that the distances $|a_n-a_{n'}|$ between 
different Gaussian packets are much larger than the width of the 
Gaussian packet, so that the overlap between different 
Gaussian packets is negligible. Hence the probability density
$\rho(x,t)=|\psi(x,t)|^2$ is
\begin{equation}
 \rho(x,t)=\frac{1}{N}\sum_{n=1}^N \rho_G(x+a_n,t) ,
\end{equation}
where $\rho_G(x+a_n,t)=|\psi_G(x+a_n,t)|^2$. Therefore (\ref{Pfinal})
gives
\begin{equation}\label{PG}
  {\cal P}(t)=\frac{v}{N}\sum_{n=1}^N \rho_G(a_n,t) 
  =\frac{v}{N}\sum_{n=1}^N \rho_G(a_n-vt,0),
\end{equation}
where in the second equality we used (\ref{shift0}) with 
$x=a_n$. This distribution has $N$ peaks at the times $t_n=a_n/v$,
for $n=1,\ldots,N$.
Since all these peaks have the same height $v\rho_G(0,0)/N$,
the distribution does not exhibit exponential time-damping.

Now let us briefly discuss why the exponential damping has been 
obtained in \cite{das-struyve}. In that work, the interpretation 
of the projected Hamiltonian $\overline{H}$ was different 
from our interpretation in Sec.~\ref{SECevol}. Basically they assumed
a Dirichlet boundary condition at $x=0$, which made $\overline{H}$
hermitian and implied a reflection of the wave packet impinging 
from the left. Hermiticity of $\overline{H}$ implies unitarity 
of $e^{-i\overline{H}t}$. Hence in this case there is no 
time-dependent normalization factor in (\ref{barcond2}),
the presence of which, as we have seen, is essential  
for the cancellation of the exponential time-damping in (\ref{Pdens}).

\section{Conclusion}
\label{SECconcl}

In this paper we have further developed the general theory of quantum time distributions introduced in \cite{jur-nik}
and applied it to arrival time distribution.
In this study, we have realized that the seemingly 
hermitian projected Hamiltonian $\overline{H}$ is in fact non-hermitian, which implies that 
the denominator in (\ref{barcond2}) is time dependent. This time dependent normalization factor 
turns out to play a crucial role for the time arrival distribution, because it cancels 
the exponential time-damping in (\ref{Pdens}). In the approximation in which the distortion of the wave 
packet (moving with velocity $v$ towards the detector at $x=0$) is neglected, we have found 
that the arrival time distribution is given by the simple formula
${\cal P}(t)=v\rho(0,t)$, where $\rho(x,t)=|\psi(x,t)|^2$ is the probability density of the wave packet.  

\section*{Acknowledgments}

The work of T.J. is supported by Croatian Science
Foundation Project No. IP-2020-02-9614.
H.N. is grateful to S. Das, D. Jurman and W. Struyve for discussions. 

\appendix

\section{Derivation of Eq.~(\ref{ee6gen}) and some related identities}
\label{APP}

Starting with the definition 
\begin{equation}
V=\bar{\pi} e^{-iH\delta t}, 
\end{equation}
the goal is to find $V^k$, in the $k\to\infty$, $\delta t\to 0$ limit, so that $k\delta t=t$ is finite.
Here $V^k$ acts on an arbitrary state $|\psi\rangle\in \mathcal{D}(H)$,
where $\mathcal{D}(H)$ is the domain of the operator $H$ and is a dense subset of the Hilbert space ${\cal H}$.
For that purpose, we can keep only terms which are proportional to 
$(\delta t)^0$ and $(\delta t)^1$, and neglect all the higher order terms proportional 
to $(\delta t)^n\equiv \delta t^n$ for $n\geq 2$. 
The case $k=1$ is trivial
\begin{equation}
 V=\bar{\pi}-i\bar{\pi}H\delta t +\mathcal{O}(\delta t^2) .
\end{equation}
The case $k=2$ is a bit more involved
\begin{eqnarray}
V^2 &=& [\bar{\pi}-i\bar{\pi}H\delta t +\mathcal{O}(\delta t^2)] V
\nonumber \\
&\stackrel{\bar{\pi}V=V}{=}& V-i\bar{\pi}H\bar{\pi}V\delta t +\mathcal{O}(\delta t^2)
\nonumber \\
&\stackrel{\bar{\pi}H\bar{\pi}=\overline{H}}{=}& V-i\overline{H}V\delta t +\mathcal{O}(\delta t^2) ,
\end{eqnarray}
after which the case $k=3$ is simple
\begin{equation}
 V^3=V^2V=V^2-i\overline{H}V^2\delta t +\mathcal{O}(\delta t^2) .
\end{equation}
By induction, we see that in general we have
\begin{equation}
V^{k+1}=V^k-i\overline{H}V^k\delta t+\mathcal{O}(\delta t^2) .
\end{equation}
Ignoring the $\mathcal{O}(\delta t^2)$, the last expression can be rearranged as
\begin{equation}\label{omjer}
\frac{V^{k+1}-V^{k}}{\delta t}=-i\overline{H}V^k ,
\end{equation}
which, in the limit $k\to\infty$, $\delta t\to 0$, with $k\delta t=t$ finite, boils down to a differential equation
for $V^k\equiv V(k\delta t)=V(t)$
\begin{equation}
\frac{d V(t)}{dt}=-i\overline{H}V(t) .
\end{equation}
The solution of the differential equation is
\begin{equation}\label{appV(t)}
V(t)=e^{-i\overline{H}t}V(0) ,
\end{equation}
where $V(0)$ is defined by $V(0)=\displaystyle\lim_{\delta t\to 0}V=\bar{\pi}$,
which proves \eqref{ee6gen}.

Since $V(t)$ is a bounded operator, the result (\ref{appV(t)}) is valid for arbitrary states $|\psi\rangle\in{\cal H}$.
But on the projected Hilbert space $\bar{\pi}{\cal H}=\overline{\cal H}$, namely on the space 
of states obeying $|\bar{\psi}\rangle=\bar{\pi}|\bar{\psi}\rangle$, 
some additional relations are valid. First, (\ref {appV(t)}) can be simplified as
\begin{equation}\label{appV(t)2}
V(t)=e^{-i\overline{H}t}, \;\;{\rm on}\; \overline{\cal H}.
\end{equation}
Second, we have
\begin{equation}\label{magic}
 \bar{\pi}H\bar{\pi}=\bar{\pi}H, \;\;{\rm on}\; \overline{\cal H}\cap {\cal D}(H),
\end{equation}
so (\ref {appV(t)2}) can also be written as
\begin{equation}\label{appV(t)3}
V(t)=e^{-i\bar{\pi}Ht}, \;\;{\rm on}\; \overline{\cal H} .
\end{equation}

Finally, using the result that (\ref{magic}) is valid on $\overline{\cal H} \cap {\cal D}(H)$, 
it is easy to see that (\ref{appV(t)}) on the full ${\cal H}$ can also be written as 
\begin{equation}\label{appV(t)4}
V(t)=e^{-i\bar{\pi}Ht}\bar{\pi} ,
\end{equation}
which generalizes (\ref{appV(t)3}) valid only on $\overline{\cal H}$.
An alternative way to obtain (\ref{appV(t)4}) is to notice that 
$\overline{H}^n\bar{\pi}=(\bar{\pi}H)^n\bar{\pi}$, which implies
\begin{eqnarray}
 e^{-i\overline{H}t}\bar{\pi} &=& \sum_n \frac{(-it)^n}{n!} \overline{H}^n\bar{\pi}
\nonumber \\
&=& \sum_n \frac{(-it)^n}{n!} (\bar{\pi}H)^n\bar{\pi} 
\nonumber \\
&=& e^{-i\bar{\pi}Ht}\bar{\pi} .
\end{eqnarray}

\section{Probability that detector will not click at $k$ subsequent 
times and the Wigner formula}
\label{APP-wigner}

In Eq.~(\ref{barcond}), which we write here as
\begin{equation}\label{barcond-app}
 |\bar{\psi}_c(t_k)\rangle \equiv |\psi_k\rangle = \frac{V^k|\psi_0\rangle}{||V^k|\psi_0\rangle||}  ,
\end{equation}
the normalization factor 
in the denominator has a direct physical interpretation;
its square $||V^k|\psi_0\rangle||^2$
{\em is the probability that the detector will not click at all $k$ subsequent times
$t_1,\ldots,t_k$}.
Here we prove this explicitly, by a method that can be viewed as a 
derivation of a special case of the Wigner formula \cite{wigner,laloe}.
We also use this to find a direct relation between this denominator and the exponential factor in (\ref{Pdens}).

Consider first the case $k=1$. The initial state $|\psi_0\rangle$ first evolves unitary from $t_0=0$ to 
$t_1=\delta t$ with the operator $e^{-iH\delta t}$. Then a measurement by the detector is performed, 
so the probability that the detector will not click at the time $t_1$ is 
$\bar{P}(1)=\langle \psi_0| e^{iH\delta t} \bar{\pi} e^{-iH\delta t}|\psi_0\rangle$.
We write this as
\begin{equation}\label{wig1}
 \bar{P}(1)=\bar{p}(1|0)=||V|\psi_0\rangle||^2,
\end{equation}
where $V\equiv \bar{\pi} e^{-iH\delta t}$ as before and $\bar{p}(1|0)$ denotes the conditional probability that 
the detector will not click at time $t_1$ given that it has not clicked at time $t_0$.

Now consider the case $k=2$. Similarly to (\ref{wig1}),
the probability that the detector will not click at times $t_1$ and $t_2$ is
\begin{equation}\label{wig2}
 \bar{P}(1,2)=\bar{p}(2|1) \bar{p}(1|0)=||V|\psi_1\rangle||^2 \; ||V|\psi_0\rangle||^2 .
\end{equation}
From (\ref{barcond-app}) we see that $|\psi_1\rangle=V|\psi_0\rangle/||V|\psi_0\rangle||$, so inserting this
into (\ref{wig2}) we get
\begin{equation}\label{wig2'}
 \bar{P}(1,2)= \frac{||V^2|\psi_0\rangle||^2}{||V|\psi_0\rangle||^2}  ||V|\psi_0\rangle||^2 
= ||V^2|\psi_0\rangle||^2 .
\end{equation}
Likewise, for $k=3$ we have
\begin{eqnarray}\label{wig3}
 \bar{P}(1,2,3) &=& \bar{p}(3|2) \bar{p}(2|1) \bar{p}(1|0)= \bar{p}(3|2) \bar{P}(1,2)
\nonumber \\ 
&=& ||V|\psi_2\rangle||^2 \; ||V^2|\psi_0\rangle||^2,
\end{eqnarray}
so inserting $|\psi_2\rangle=V^2|\psi_0\rangle/||V^2|\psi_0\rangle||$ we get
\begin{equation}\label{wig3'}
 \bar{P}(1,2,3)=||V^3|\psi_0\rangle||^2 . 
\end{equation}
Then by induction, for arbitrary $k$ we have 
\begin{eqnarray}\label{wigk}
 \bar{P}(1,2,\ldots, k) &=& \bar{p}(k|k-1)\cdots \bar{p}(2|1) \bar{p}(1|0)
\nonumber \\
&=& \bar{p}(k|k-1) \bar{P}(1,2,\ldots,k-1)
\nonumber \\ 
&=& ||V^k|\psi_0\rangle||^2.
\end{eqnarray}

Our final result $(\ref{wigk})$ is nothing but a special case of Wigner formula \cite{wigner,laloe},
so for the sake of completeness let us write one version of the 
general Wigner formula. 
If at each time $t_i$ the system exhibits a collapse defined by some projector $\pi_i$, 
then the probability of the whole series of collapses is 
\begin{equation}\label{wigk'}
 P(1,2,\ldots, k)=||\pi_kU_k\cdots \pi_2U_2\pi_1U_1|\psi_0\rangle||^2,
\end{equation}
where $U_i$ is the operator of unitary evolution from time $t_{i-1}$ to time $t_i$. 
Eq.~(\ref{wigk'}) is the Wigner formula, which can be proved \cite{laloe} 
by a straightforward generalization of the proof of (\ref{wigk}) above.

Not let us see how is the probability (\ref{wigk}) related to the exponential factor in (\ref{Pdens}).
The first line of (\ref{wigk}) can be written as
\begin{eqnarray}\label{wig_exp}
 \bar{P}(1,2,\ldots, k) &=& \prod_{i=1}^k \bar{p}(i|i-1)=\prod_{i=1}^k e^{{\rm ln}\; \bar{p}(i|i-1)}
\nonumber \\
&=& \exp\left( \sum_{i=1}^k {\rm ln}\; \bar{p}(i|i-1) \right) .
\end{eqnarray}
We recall that $\bar{p}(i|i-1)$ is the probability that the detector will not click at time $t_i$,
given that it has not clicked at time $t_{i-1}$. Hence we can write 
$\bar{p}(i|i-1)=1-p(i|i-1)$, where $p(i|i-1)$ is the probability that the detector {\em will} click at time $t_i$,
given that it has not clicked at time $t_{i-1}$.
When $t_i-t_{i-1}\equiv \delta t$ goes to zero, then $p(i|i-1)$ also goes to zero.
Hence, since we are interested in a non-zero but small $\delta t$, we have $p(i|i-1)\ll 1$ so
\begin{eqnarray}
  {\rm ln}\; \bar{p}(i|i-1) &=& {\rm ln}[1-p(i|i-1)] 
\nonumber \\
& \simeq & -p(i|i-1) = -p(t_i), 
\end{eqnarray}
where in the last equality we have used the fact that $p(i|i-1)$ is just a different notation for the conditional probability
$p(t_i)$ in (\ref{p}).
Hence (\ref{wig_exp}) can be approximated with
\begin{eqnarray}\label{wig_exp3}
 \bar{P}(1,2,\ldots, k) & \simeq & \exp\left( -\sum_{i=1}^k p(t_i) \right) 
\nonumber \\
& \simeq &  e^{-\int_0^t dt' w(t')} ,
\end{eqnarray}
where in the second line we have approximated the sum with the integral 
(which is justified because $\delta t$ is small \cite{jur-nik})
and $w(t)$ is defined as in (\ref{w}).

The upshot of all this is that, in the limit $\delta t\to 0$,
(\ref{wigk}) and (\ref{wig_exp3}) imply a useful formula
\begin{equation}\label{wig-up}
||V^k|\psi_0\rangle||^2 = e^{-\int_0^t dt' w(t')} .
\end{equation}
One immediate application of this formula is to combine (\ref{barcond}) with (\ref{ee6}) to write 
\begin{equation}\label{barcond-app2}
|\bar{\psi}_c(t)\rangle=\frac{e^{-i\overline{H}t}|\psi_0\rangle}{e^{-\int_0^t dt' w(t')/2}} .
\end{equation} 

\section*{Statements and Declarations}

Data sharing not applicable to this article as no datasets were generated or analyzed during the current study 
and article describes entirely theoretical research.



\begin{thebibliography}{99}

\bibitem{jur-nik}
D. Jurman and H. Nikoli\'c, Phys. Lett. A {\bf 396}, 127247 (2021); arXiv:2010.07575.

\bibitem{muga1}
J.G. Muga, R. Sala Mayato, and I.L. Egesquiza (eds), {\it Time in Quantum Mechanics - Vol. 1}
(Springer-Werlag, Berlin, 2008).
\bibitem{muga2}
J.G. Muga, A. Ruschhaupt, and A. del Campo (eds), {\it Time in Quantum Mechanics - Vol. 2}
(Springer-Werlag, Berlin, 2009).

\bibitem{marchewka}
A. Marchewka and Z. Schuss, Phys. Rev. A {\bf 65}, 042112 (2002).
\bibitem{wlodarz}
J.J. Wlodarz, Phys. Rev. A {\bf 65}, 044103 (2002).
%
\bibitem{V1.10}
I.L. Egusquiza, J.G. Muga, and A.D. Baute, 
Chapter 10
in \cite{muga1}.
\bibitem{V2.4}
A. Ruschhaupt, J.G. Muga, and G.C. Hegerfeldt, 
Chapter 4
in \cite{muga2}.
\bibitem{rovelli}
N. Grot, C. Rovelli, and R.S. Tate, Phys. Rev. A {\bf 54}, 4679 (1996); quant-ph/9603021.
\bibitem{delgado}
V. Delgado and J.G. Muga, Phys. Rev. A {\bf 56}, 3425 (1997); quant-ph/9704010.
\bibitem{galapon2}
E.A. Galapon, R.F. Caballar, and R.T. Bahague Jr, Phys. Rev. Let. {\bf 93}, 180406 (2004); quant-ph/0302036.
\bibitem{anastopoulos2}
C. Anastopoulos and N. Savvidou,  J. Math. Phys. {\bf 47}, 122106 (2006); quant-ph/0509020.
\bibitem{halliwell1}
J.J. Halliwell and J.M. Yearsley, Phys. Lett. A {\bf 374}, 154 (2009); arXiv:0903.1958.
\bibitem{anastopoulos}
C. Anastopoulos and N. Savvidou, Phys. Rev. A {\bf 86}, 012111 (2012);  arXiv:1205.2781.
\bibitem{vona}
N. Vona, G. Hinrichs, and D. D\"urr, Phys. Rev. Lett. {\bf 111}, 220404 (2013); arXiv:1307.4366.
\bibitem{dhar}
S. Dhar, S. Dasgupta, and A. Dhar, J. Phys. A: Math. Theor. {\bf 48} 115304 (2015); arXiv:1312.5923.
\bibitem{halliwell2}
J.J. Halliwell, J. Evaeus, J. London, and Y. Malik, Phys. Lett. A {\bf 379}, 2445 (2015); arXiv:1504.02509.
\bibitem{galapon}
E.A. Galapon, J. Jaykel, and P. Magadan, Ann. Phys. {\bf 397}, 278 (2018); arXiv:1804.03344.
%
\bibitem{leavens}
C.R. Leavens, 
Chapter 5
in \cite{muga1}.
\bibitem{durrtime1}
S. Das and D. D\"urr, Scientific Reports {\bf 9}, 2242 (2019); arXiv:1802.07141.
\bibitem{durrtime2}
S. Das, M. N\"oth, and D. D\"urr, Phys. Rev. A {\bf 99}, 052124 (2019); arXiv:1901.08672.
%
\bibitem{maccone}
L. Maccone and K. Sacha, Phys. Rev. Lett. {\bf 124}, 110402 (2020); arXiv:1810.12869.
%
\bibitem{muga-physrep}
J.G. Muga and C.R. Leavens, Phys. Rep. {\bf 338}, 353 (2000).
%

\bibitem{allcock}
G.R. Allcock, Ann. Phys. {\bf 53}, 253 (1969);
G.R. Allcock, Ann. Phys. {\bf 53}, 286 (1969);
G.R. Allcock, Ann. Phys. {\bf 53}, 311 (1969).
\bibitem{echanobe}
J. Echanobe, A. del Campo, and J.G. Muga, Phys. Rev. A {\bf 77}, 032112 (2008); arXiv:0712.0670.

\bibitem{muga99}
J.G. Muga, J.P. Palao, and C.R. Leavens, Phys. Lett. A {\bf 253}, 21 (1999). 
\bibitem{halliwell08}
J.J. Halliwell, Phys. Rev. A {\bf 77}, 062103 (2008); arXiv:0801.4308.
\bibitem{halliwell10}
J.J. Halliwell and J.M. Yearsley, J. Phys. A: Math. Theor. {\bf 43}, 445303 (2010); arXiv:1006.4788.
\bibitem{kiukas}
J. Kiukas, A. Ruschhaupt, P.O. Schmidt, and R.F. Werner, J. Phys. A: Math. Theor. {\bf 45}, 185301 (2012); arXiv:1109.5087.


\bibitem{anastopoulos3}
C. Anastopoulos and N. Savvidou, J. Math. Phys. {\bf 60}, 032301 (2019); arXiv:1807.06533.

\bibitem{das-struyve}
S. Das and W. Struyve, Phys. Rev. A {\bf 104}, 042214 (2021); arXiv:2105.14744.

\bibitem{wigner}
E.P. Wigner, Chapter II.4 in
J.A. Wheeler and W.H. Zurek (eds.), {\it Quantum Theory and Measurement} (Princeton University Press, Princeton, 1983). 
\bibitem{laloe}
F. Lalo\"e, {\it Do We Really Understand Quantum Mechanics?} (Cambridge University Press, Cambridge, 2019).

\bibitem{gieres}
F. Gieres, Rep. Prog. Phys. {\bf 63}, 1893 (2000); quant-ph/9907069.

\bibitem{gitman}
D.M. Gitman, I.V. Tyutin, and B.L. Voronov, {\it Self-adjoint Extensions in Quantum Mechanics}
(Springer, New York, 2012).
\bibitem{juric}
T. Juri\'c, Universe {\bf 8}, 129 (2022); arXiv:2103.01080.
\bibitem{reed}
M. Reed and B. Simon, {\it Methods of Modern Mathematical Physics, Vols. 1 and 2}
(Academic Press, 1978).

\bibitem{cohen}
C. Cohen-Tannoudji, B. Diu, and F. Lalo\"e, {\it Quantum Mechanics, Vol. I} 
(WILEY, Weinheim, 2020).



\end{thebibliography}
\end{document}